\documentclass[10pt,twocolumn]{IEEEtran}

\usepackage{verbatim,epsfig,rotating,setspace,latexsym,amsmath,epsf,amssymb}
\usepackage{cite,psfrag,graphicx,amsmath, amssymb}
\usepackage{epsfig,rotating,setspace,latexsym,amsmath,epsf,amssymb,bm,amsbsy,float}

\usepackage{cite,graphicx,color,subfigure,here}

\begin{document}

\pagestyle{empty}
\title{A New Upper Bound on the Capacity of a Class of Primitive Relay Channels\thanks{This work was supported by
NSF Grants CCF $04$-$47613$, CCF $05$-$14846$, CNS $07$-$16311$
and CCF 07-29127. }}

\author{
\authorblockN{Ravi Tandon \qquad Sennur Ulukus }\\
\authorblockA{Department of Electrical and Computer Engineering \\
University of Maryland, College Park, MD 20742 \\
{\textit{ravit@umd.edu}} \qquad {\textit{ulukus@umd.edu}}} }

\newcounter{mytempeqncnt}

\newcommand{\av}{\mathbf{a}}
\newcommand{\bv}{\mathbf{b}}
\newcommand{\cv}{\mathbf{c}}
\newcommand{\dv}{\mathbf{d}}
\newcommand{\ev}{\mathbf{e}}
\newcommand{\fv}{\mathbf{f}}
\newcommand{\gv}{\mathbf{g}}
\newcommand{\hv}{\mathbf{h}}
\newcommand{\iv}{\mathbf{i}}
\newcommand{\jv}{\mathbf{j}}
\newcommand{\kv}{\mathbf{k}}
\newcommand{\lv}{\mathbf{l}}
\newcommand{\mv}{\mathbf{m}}
\newcommand{\nv}{\mathbf{n}}
\newcommand{\ov}{\mathbf{o}}
\newcommand{\pv}{\mathbf{p}}
\newcommand{\qv}{\mathbf{q}}
\newcommand{\rv}{\mathbf{r}}
\newcommand{\sv}{\mathbf{s}}
\newcommand{\tv}{\mathbf{t}}
\newcommand{\uv}{\mathbf{u}}
\newcommand{\vv}{\mathbf{v}}
\newcommand{\xv}{\mathbf{x}}
\newcommand{\wv}{\mathbf{w}}
\newcommand{\yv}{\mathbf{y}}
\newcommand{\zv}{\mathbf{z}}

\newcommand{\Amat}{\mathbf{A}}
\newcommand{\Bmat}{\mathbf{B}}
\newcommand{\Cmat}{\mathbf{C}}
\newcommand{\Dmat}{\mathbf{D}}
\newcommand{\Emat}{\mathbf{E}}
\newcommand{\Fmat}{\mathbf{F}}
\newcommand{\Gmat}{\mathbf{G}}
\newcommand{\Hmat}{\mathbf{H}}
\newcommand{\Imat}{\mathbf{I}}
\newcommand{\Jmat}{\mathbf{J}}
\newcommand{\Kmat}{\mathbf{K}}
\newcommand{\Lmat}{\mathbf{L}}
\newcommand{\Mmat}{\mathbf{M}}
\newcommand{\Nmat}{\mathbf{N}}
\newcommand{\Omat}{\mathbf{O}}
\newcommand{\Pmat}{\mathbf{P}}
\newcommand{\Qmat}{\mathbf{Q}}
\newcommand{\Rmat}{\mathbf{R}}
\newcommand{\Smat}{\mathbf{S}}
\newcommand{\Tmat}{\mathbf{T}}
\newcommand{\Umat}{\mathbf{U}}
\newcommand{\Vmat}{\mathbf{V}}
\newcommand{\Xmat}{\mathbf{X}}
\newcommand{\Wmat}{\mathbf{W}}
\newcommand{\Ymat}{\mathbf{Y}}
\newcommand{\Zmat}{\mathbf{Z}}

\newcommand{\Zeromat}{\mathbf{0}}

\newcommand{\lam}{\bm{\lambda}}
\newcommand{\thetav}{\bm{\theta}}
\newcommand{\varthetav}{\bm{\vartheta}}

\newcommand{\Lam}{\bm{\Lambda}}
\newcommand{\Ximat}{\bm{\Xi}}
\newcommand{\Phimat}{\bm{\Phi}}
\newcommand{\Sig}{\bm{\Sigma}}
\newcommand{\Del}{\bm{\Delta}}
\newcommand{\Ccal}{\cal C}
\newcommand{\tr}{\text{tr}}

\newtheorem{Theo}{Theorem}
\newtheorem{Lem}{Lemma}
\newtheorem{Cor}{Corollary}
\newtheorem{Def}{Definition}

\maketitle \thispagestyle{empty}

\begin{abstract}
We obtain a new upper bound on the capacity of a class of discrete
memoryless relay channels. For this class of relay channels, the
relay observes an i.i.d. sequence $T$, which is independent of the
channel input $X$. The channel is described by a set of
probability transition functions $p(y|x,t)$ for all $(x,t,y)\in
\mathcal{X}\times \mathcal{T}\times \mathcal{Y}$. Furthermore, a
noiseless link of finite capacity $R_{0}$ exists from the relay to
the receiver. Although the capacity for these channels is not
known in general, the capacity of a subclass of these channels,
namely when $T=g(X,Y)$, for some deterministic function $g$, was
obtained in \cite{KimRelay:2008} and it was shown to be equal to
the cut-set bound. Another instance where the capacity was
obtained was in \cite{YuRelay:2007}, where the channel output $Y$
can be written as $Y=X\oplus Z$, where $\oplus$ denotes modulo-$m$
addition, $Z$ is independent of $X$,
$|\mathcal{X}|=|\mathcal{Y}|=m$, and $T$ is some stochastic
function of $Z$. The compress-and-forward (CAF) achievability
scheme \cite{Cover:1979} was shown to be capacity achieving in
both cases.

Using our upper bound we recover the capacity results of
\cite{KimRelay:2008} and \cite{YuRelay:2007}. We also obtain the
capacity of a class of channels which does not fall into either of
the classes studied in \cite{KimRelay:2008} and
\cite{YuRelay:2007}. For this class of channels, CAF scheme is
shown to be optimal but capacity is strictly less than the cut-set
bound for certain values of $R_{0}$. We further illustrate the
usefulness of our bound by evaluating it for a particular relay
channel with binary multiplicative states and binary additive
noise for which the channel is given as $Y=TX+N$. We show that our
upper bound is strictly better than the cut-set upper bound for
certain values of $R_{0}$ but it lies strictly above the rates
yielded by the CAF achievability scheme.
\end{abstract}

\section{Introduction}
The relay channel is one of the simplest, yet arguably among the
least understood multi-user channels in information theory. A
special class of discrete memoryless relay channel is the
primitive relay channel \cite{KimRelay:2008}. For this class, the
channel is defined by a channel input $X$, a channel output $Y$
and a relay output $T$, and a set of probability functions
$p(y,t|x)$ for all $x\in \mathcal{X}$. In this setting, the relay
does not have an explicit coded input for the channel. Moreover,
it is also assumed that there is an orthogonal link of finite
capacity $R_{0}$, from the relay to the receiver. Zhang
\cite{Zhang:1988} considered this relay channel and obtained a
partial converse for a degraded case. For a comprehensive survey
on related work on primitive relay channels, see
\cite{KimAllerton:2007}.

Recently, Kim \cite{KimRelay:2008} established the capacity of a
class of semi-deterministic primitive relay channels, for which
the relay output $T$ can be expressed as a deterministic function
of the channel input $X$ and the channel output $Y$, i.e.,
$T=g(Y,X)$. The cut-set upper bound \cite{Cover:book} was shown to
be the capacity through an algebraic reduction of the
compress-and-forward (CAF) achievable rate \cite{Cover:1979} to
the cut-set upper bound. This was the first instance where the CAF
achievability scheme was shown to be capacity achieving for any
relay channel.

In this paper, we consider a subclass of the primitive relay
channel. In this subclass, the relay observes an i.i.d. sequence
$T$ which is independent of the channel input $X$ and the channel
output $Y$ is given by the set of probability transition functions
$p(y|x,t)$ for all $(x,t,y)\in
\mathcal{X}\times\mathcal{T}\times\mathcal{Y}$. Alternatively,
this channel can be interpreted as a state dependent discrete
memoryless channel with rate-limited state information available
at the receiver (Figure $1$). This channel was also studied in
\cite{HeegardElGamal:1983} with various modifications regarding
the rate-limited knowledge of the channel state $T$ at the
transmitter and the receiver. A CAF achievability scheme for this
state dependent channel was given by Ahlswede and Han in
\cite{AhlswedeHan:1983} and it was conjectured to be the capacity
for this class of channels. In fact, the same achievable rates for
this channel were obtained in \cite{HeegardElGamal:1983} and can
also be obtained via Theorem $6$ of \cite{Cover:1979}.

It follows from the result of \cite{KimRelay:2008} that this
conjecture is true for the subclass when the state $T$ can be
expressed as a deterministic function of $X$ and $Y$, i.e.,
$T=g(X,Y)$. An example of such a channel is the case when $X$, $T$
and $Y$ are all binary, $T\sim\text{Ber}(\delta)$ and independent
of $X$, and the channel is given by $Y=X\oplus T$, where $\oplus$
denotes modulo-$2$ addition. Note that, in this case, $T$ is a
deterministic function of $X$ and $Y$, since $T=X\oplus Y$. A
capacity result following up on the aforementioned modulo-additive
noise channel was obtained in \cite{YuRelay:2007}, where it was
assumed that the receiver observes $Y=X\oplus Z$ and the relay
observes a noisy version of the forward noise, i.e., $T=Z\oplus
\tilde{Z}$. Clearly, if $\tilde{Z}=0$, then this channel reduces
to the class studied in \cite{KimRelay:2008}. However, when
$\tilde{Z}\neq 0$, $T$ cannot be written as a deterministic
function of $X$ and $Y$, and this modulo-additive class lies
outside of the class of channels considered in
\cite{KimRelay:2008}. By proving a converse, it was shown in
\cite{YuRelay:2007} that CAF scheme is capacity achieving for this
modulo-additive case. The remarkable fact was that the capacity
was shown to be strictly less than the cut-set upper bound for
certain values of $R_{0}$. However, it is worth noting that the
converse proved in \cite{YuRelay:2007} relied heavily on the
modulo-additive nature of the forward channel.

In this paper, we obtain a new upper bound on the capacity of the
state-dependent discrete memoryless channel, where the states are
i.i.d. and the state information is available to the receiver
through a noiseless link of finite capacity $R_{0}$. Our upper
bound serves a dual purpose. Firstly, using our upper bound, we
recover the capacity results obtained in \cite{KimRelay:2008} for
the case where $T=g(X,Y)$ and the capacity result obtained in
\cite{YuRelay:2007} for the modulo-additive noise case. Secondly,
we confirm the validity of the conjecture due to Ahlswede-Han
\cite{AhlswedeHan:1983} for another class of channels which does
not fall into any of the cases considered in \cite{KimRelay:2008}
and \cite{YuRelay:2007}.
\begin{figure}[t]
  \centering
  \centerline{\epsfig{figure=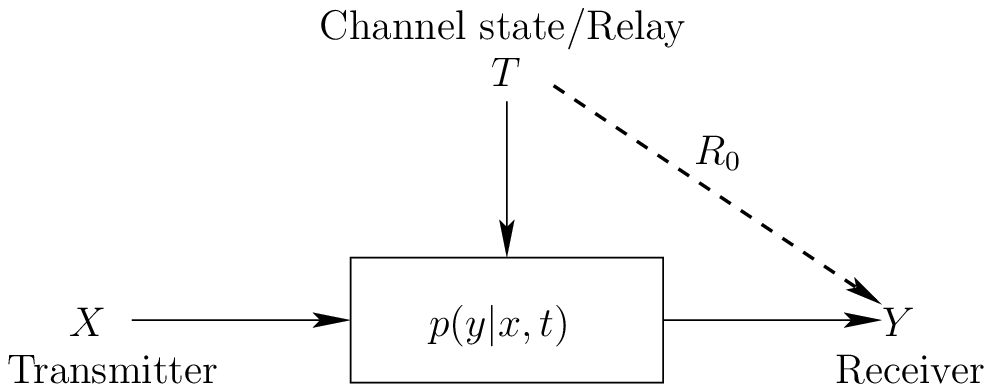,width=7cm}}
  \vspace{0.1in}
  { Figure $1$: Channel with rate-limited state information.}\medskip
  \vspace{-0.3in}
\end{figure}

To further illustrate the application of our upper bound, we
consider a channel where $X$, $T$, $N$ are binary and $Y$ is
ternary and the channel is given by $Y=TX+N$, i.e., when the state
sequence is binary and multiplicative and there is additive binary
noise at the receiver. This channel can be interpreted as the
discrete analogue of a fast fading channel with fade information
available in a rate-limited fashion at the receiver. This channel
does not fall into any class for which capacity is known. We
evaluate our upper bound for this channel and show that it is
strictly less than the cut-set bound for certain values of $R_{0}$
although our upper bound is strictly larger than the rates yielded
by the CAF scheme. \vspace{-0.05in}

\section{Relay Channel Model}
We consider a relay channel with finite input alphabet
$\mathcal{X}$, finite output alphabet $\mathcal{Y}$ and finite
relay output alphabet $\mathcal{T}$. Moreover, the relay observes
an i.i.d. state sequence $T^{n}\in \mathcal{T}^{n}$ with some
given probability distribution $p(t)$. The relay channel is
described by the set of transition probabilities $p(y|x,t)$ which
are defined for all $(x,t,y)\in
\mathcal{X}\times\mathcal{T}\times\mathcal{Y}$. Furthermore, there
is a finite-capacity noiseless link of capacity $R_{0}$ from the
relay to the receiver. This relay channel can also be thought of
as a state-dependent single-user channel with rate-limited state
information available at the receiver (see Figure $1$).

An $(n,M,P_{e})$ code for this relay channel consists of the set
of integers $\mathcal{M}=\{1,2,\ldots,M\}$ and the following:
\begin{align}
f_{t}&:\mathcal{M}\rightarrow \mathcal{X}^{n}\nonumber\\
f_{r}&:\mathcal{T}^{n}\rightarrow \{1,2,\ldots,L\}\nonumber\\
\phi&:\mathcal{Y}^{n}\times \{1,2,\ldots,L\}\rightarrow
\mathcal{M}\label{Model1}
\end{align}
where $f_{t}$ is the transmitter encoding function, $f_{r}$ is the
relay encoding function and $g$ is the decoding function.
Furthermore, as the relay to receiver link is of limited capacity
$R_{0}$, we have
\begin{align}
L\leq 2^{nR_{0}}\label{Model2}
\end{align}
For a distribution $p(w)$ on $\mathcal{M}$, the joint probability
distribution on $\mathcal{M}\times
\mathcal{X}^{n}\times\mathcal{T}^{n}\times\mathcal{Y}^{n}$ is
given as
\begin{align}
p(w,x^{n},t^{n},y^{n})&=p(w)p(x^{n}|w)\prod_{i=1}^{n}p(t_{i})\prod_{i=1}^{n}p(y_{i}|x_{i},t_{i})\label{Model3}
\end{align}
For a uniform distribution $p(w)$ on $\mathcal{M}$, the average
probability of error is given as,
$P_{e}=\mbox{Pr}(\phi(Y^{n},f_{r}(T^{n}))\neq W)\label{Model3}$. A
rate $R$ is achievable if for any $\epsilon >0$ and all $n$
sufficiently large, there exists an $(n,M,P_{e})$ code such that
$P_{e}\leq \epsilon$ and $M\geq 2^{nR}$. The capacity of the relay
channel is the supremum of the set of all achievable rates.

\section{A New Upper Bound on the Capacity}
We will denote by $U^{n}$ as the output of the finite capacity
link $R_{0}$, i.e., $U^{n}=f_{r}(T^{n})$. We will now obtain an
upper bound on the rate as follows,
\begin{align}
nR&=H(W)\label{ub1}\\
&=I(W;Y^{n},U^{n})+H(W|Y^{n},U^{n})\label{ub2}\\
&\leq I(W;Y^{n},U^{n})+n\epsilon_{n}\label{ub3}\\
&\leq I(X^{n};Y^{n},U^{n})+n\epsilon_{n}\label{ub4}\\
&=I(X^{n};Y^{n}|U^{n})+n\epsilon_{n}\label{ub5}\\
&=\sum_{i=1}^{n}I(X^{n};Y_{i}|U^{n},Y^{i-1})+n\epsilon_{n}\label{ub6}\\
&=\sum_{i=1}^{n}\Big[H(Y_{i}|U^{n},Y^{i-1})-H(Y_{i}|U^{n},Y^{i-1},X^{n})\Big]+n\epsilon_{n}\label{ub7}\\
&\leq
\sum_{i=1}^{n}\Big[H(Y_{i})-H(Y_{i}|U^{n},Y^{i-1},X^{n})\Big]+n\epsilon_{n}\label{ub8}\\
&\leq \sum_{i=1}^{n}\Big[H(Y_{i})-H(Y_{i}|U^{n},T^{i-1},Y^{i-1},X^{n})\Big]+n\epsilon_{n}\label{ub9}\\
&= \sum_{i=1}^{n}\Big[H(Y_{i})-H(Y_{i}|U^{n},T^{i-1},X^{n})\Big]+n\epsilon_{n}\label{ub10}\\
&= \sum_{i=1}^{n}\Big[H(Y_{i})-H(Y_{i}|U^{n},T^{i-1},X_{i})\Big]+n\epsilon_{n}\label{ub11}\\
&= \sum_{i=1}^{n}I(X_{i},U^{n},T^{i-1};Y_{i})+n\epsilon_{n}\label{ub12}\\
&= \sum_{i=1}^{n}I(X_{i},V_{i};Y_{i})+n\epsilon_{n}\label{ub13}\\
&= nI(X,V;Y)+n\epsilon_{n}\label{ub14}
\end{align}
where (\ref{ub3}) follows by Fano's inequality \cite{Cover:book},
(\ref{ub4}) follows from the data processing inequality,
(\ref{ub5}) follows from the fact that $X^{n}$ is independent of
$T^{n}$ and is hence independent of $U^{n}$, (\ref{ub8}) follows
from the fact that conditioning reduces entropy and hence we upper
bound by dropping $(U^{n},Y^{i-1})$ from the first term. Next,
(\ref{ub9}) follows by adding $T^{i-1}$ in the conditional entropy
in the second term and obtaining an upper bound,
(\ref{ub10}) follows from the memoryless property of the channel,
i.e., given $(X^{i-1},T^{i-1})$, the channel output $Y^{i-1}$ is
independent of everything else and (\ref{ub11}) follows from
the following Markov chain, $X^{n}\setminus X_{i} \rightarrow
(X_{i},U^{n},T^{i-1}) \rightarrow Y_{i} $. The proof of this
Markov chain is given at the beginning of next page. Finally, (\ref{ub13}) follows by defining $V_{i}=(U^{n},T^{i-1})$,
and we introduce a random variable $Q$, uniform on
$\{1,2,\ldots,n\}$ to define $X=(X_{i},Q)$, $Y=(Y_{i},Q)$ and
$V=(V_{i},Q)$ to arrive at (\ref{ub14}).

\begin{figure*}[t]
\setcounter{mytempeqncnt}{\value{equation}}

\setcounter{equation}{17} We obtain the Markov chain by showing
the following,
\begin{align}
\mbox{Pr}(Y_{i},X^{-i}|X_{i},U^{n},T^{i-1})&=\frac{\mbox{Pr}(Y_{i},X^{-i},X_{i},U^{n},T^{i-1})}{\mbox{Pr}(X_{i},U^{n},T^{i-1})}\label{mc2}\\
&=\frac{\sum_{t_{i}}\mbox{P}(t_{i})\mbox{Pr}(Y_{i},X^{-i},X_{i},U^{n},T^{i-1}|t_{i})}{\mbox{Pr}(X_{i},U^{n},T^{i-1})}\label{mc3}\\
&=\frac{\sum_{t_{i}}\mbox{P}(t_{i})\mbox{Pr}(X_{i},U^{n},T^{i-1}|t_{i})\mbox{Pr}(Y_{i},X^{-i}|X_{i},t_{i},U^{n},T^{i-1})}{\mbox{Pr}(X_{i},U^{n},T^{i-1})}\label{mc4}\\
&=\frac{\sum_{t_{i}}\mbox{P}(t_{i})\mbox{Pr}(X_{i},U^{n},T^{i-1}|t_{i})\mbox{Pr}(X^{-i}|X_{i},t_{i},U^{n},T^{i-1})\mbox{Pr}(Y_{i}|X_{i},t_{i},U^{n},T^{i-1},X^{-i})}{\mbox{Pr}(X_{i},U^{n},T^{i-1})}\label{mc5}\\
&=\frac{\sum_{t_{i}}\mbox{P}(t_{i})\mbox{Pr}(X_{i},U^{n},T^{i-1}|t_{i})\mbox{Pr}(X^{-i}|X_{i})\mbox{Pr}(Y_{i}|X_{i},t_{i},U^{n},T^{i-1})}{\mbox{Pr}(X_{i},U^{n},T^{i-1})}\label{mc6}\\
&=\mbox{Pr}(X^{-i}|X_{i})\frac{\sum_{t_{i}}\mbox{P}(t_{i})\mbox{Pr}(X_{i},U^{n},T^{i-1}|t_{i})\mbox{Pr}(Y_{i}|X_{i},t_{i},U^{n},T^{i-1})}{\mbox{Pr}(X_{i},U^{n},T^{i-1})}\label{mc7}\\
&=\mbox{Pr}(X^{-i}|X_{i})\sum_{t_{i}}\mbox{P}(t_{i}|X_{i},U^{n},T^{i-1})\mbox{Pr}(Y_{i}|X_{i},U^{n},T^{i-1},t_{i})\label{mc8}\\
&=\mbox{Pr}(X^{-i}|X_{i})\mbox{Pr}(Y_{i}|X_{i},U^{n},T^{i-1})\label{mc9}
\end{align}
where we have defined $X^{-i}\triangleq
(X_{1},\ldots,X_{i-1},X_{i+1},\ldots,X^{n})$.
\setcounter{equation}{25}

\hrulefill
\end{figure*}

In addition to (\ref{ub14}), we also need the following trivial
upper bound on the rate,
\begin{align}
nR&\leq I(X^{n};Y^{n},T^{n})+n\epsilon_{n}\label{ub15}\\
&= I(X^{n};Y^{n}|T^{n})+n\epsilon_{n}\label{ub16}\\
&=\sum_{i=1}^{n}I(X^{n};Y_{i}|T^{n},Y^{i-1})+n\epsilon_{n}\label{ub17}\\
&=\sum_{i=1}^{n}\Big[H(Y_{i}|T^{n},Y^{i-1})-H(Y_{i}|T^{n},Y^{i-1},X^{n})\Big]+n\epsilon_{n}\hspace{0.05in}\label{ub18}\\
&=\sum_{i=1}^{n}\Big[H(Y_{i}|T_{i})-H(Y_{i}|T^{n},Y^{i-1},X^{n})\Big]+n\epsilon_{n}\label{ub19}\\
&=\sum_{i=1}^{n}\Big[H(Y_{i}|T_{i})-H(Y_{i}|T_{i},X_{i})\Big]+n\epsilon_{n}\label{ub20}\\
&= \sum_{i=1}^{n}I(X_{i};Y_{i}|T_{i})+n\epsilon_{n}\label{ub21}\\
&=nI(X;Y|T)+n\epsilon_{n}\label{ub22}
\end{align}
where (\ref{ub15}) follows by Fano's inequality, (\ref{ub16})
follows because $X^{n}$ is independent of $T^{n}$, (\ref{ub19})
follows by dropping $(Y^{i-1},T^{n}\setminus T_{i})$ from the
conditioning in the first term, (\ref{ub20}) follows from the
memoryless property of the channel, i.e., given $(X_{i},T_{i})$,
the channel output $Y_{i}$ is independent of everything else.

We now obtain a bound on the allowable distributions of the
involved random variables. Using the fact that the side
information is limited by the rate $R_{0}$, we have that
\begin{align}
n R_{0}&\geq I(T^{n};U^{n})\label{ub23}\\
&=\sum_{i=1}^{n}I(T_{i};U^{n}|T^{i-1})\label{ub24}
\end{align}
\begin{align}
&=\sum_{i=1}^{n}I(T_{i};U^{n},T^{i-1})\label{ub25}\\
&=nI(T;V) \label{ub26}
\end{align}
where (\ref{ub25}) follows from the fact that $T_{i}$ are i.i.d.

Combining (\ref{ub14}), (\ref{ub22}) and (\ref{ub26}), we have an
upper bound on the capacity of the relay channel as
\begin{align}
\mathcal{UB}&= \sup \min \{I(X,V;Y),I(X;Y|T)\}\nonumber\\
&\hspace{0.2in}\text{s.t. }R_{0}\geq I(T;V)\nonumber\\
&\hspace{0.2in}\text{over }p(x)p(t)p(v|t)\label{UB}
\end{align}
where the supremum can be restricted over those $V$ such that
$|\mathcal{V}|\leq |\mathcal{T}|+2$.

\section{Comparison with the Cut-set Bound}
The best known upper bound for the relay channel is the cut-set
bound \cite{Cover:book}, which reduces for the relay channel in
consideration to \cite{KimRelay:2008,KimAllerton:2007}
\begin{align}
\mathcal{CS}&=\max_{p(x)} \min
\{I(X;Y)+R_{0},I(X;Y|T)\}\label{cutset}
\end{align}
On comparing with the cut-set bound, it can be observed that our
bound differs from the cut-set bound in the multiple access cut.
We will show next that our upper bound is in general smaller than
the cut-set bound.

We start by upper bounding the expression $I(X,V;Y)$ as follows,
\begin{align}
I(X,V;Y)&=I(X;Y)+I(V;Y|X)\label{K1}\\
&=I(X;Y)+H(V|X)-H(V|Y,X)\label{K2}\\
&=I(X;Y)+H(V)-H(V|Y,X)\label{K3}\\
&\leq I(X;Y)+H(V)-H(V|T,Y,X)\label{K4}\\
&=I(X;Y)+H(V)-H(V|T)\label{K5}
\end{align}
\begin{align}
&=I(X;Y)+I(T;V)\label{K6}\\
&\leq I(X;Y)+R_{0}\label{K7}
\end{align}
where (\ref{K3}) follows from the fact that $V$ is independent of
$X$, (\ref{K4}) follows from the fact that conditioning reduces
entropy, (\ref{K5}) follows from the Markov chain
$(X,Y)\rightarrow T \rightarrow V$ and (\ref{K6}) follows by using
the fact that $I(T;V)\leq R_{0}$. Using (\ref{K7}) and
(\ref{ub22}), we have the following
\begin{align}
\mathcal{UB}&\leq
\max_{p(x)}\min\{I(X;Y)+R_{0},I(X;Y|T)\}\label{K8}
\end{align}
Thus, our upper bound is in general smaller than the cut-set bound
given in (\ref{cutset}). It was shown in \cite{KimRelay:2008} that
the cut-set bound is tight for the case when $T=g(X,Y)$ and is
achieved by the CAF achievability scheme. Note the fact that for
this special subclass, the inequality in (\ref{K4}) is in fact an
equality and our bound exactly equals the cut-set bound.

\section{Recovering the Capacity of Modulo-Additive Relay Channel}
A specific modulo-additive relay channel was considered in
\cite{YuRelay:2007} for which the channel is given as,
\begin{align}
Y&=X\oplus Z\label{Yu1}\\
T&=Z\oplus \tilde{Z}\label{Yu2}
\end{align}
where $X$, $Y$, $T$, $Z$ and $\tilde{Z}$ are all binary and $Z\sim
\mbox{Ber}(\delta)$, $\tilde{Z}\sim \mbox{Ber}(\tilde{\delta})$.
Clearly this channel does not fall into the class of channels
studied in \cite{KimRelay:2008}, where $T$ can be written as a
deterministic function of $X$ and $Y$. It was shown that the
capacity of this channel is given by \cite[Theorem
$1$]{YuRelay:2007}
\begin{align}
\mathcal{C}&=\max_{p(v|t):I(T;V)\leq R_{0}} 1-H(Z|V)\label{Yu3}
\end{align}
We will show that our bound is equal to the capacity for this
class of channels. First, note that
\begin{align}
I(X,V;Y)&=H(Y)-H(Y|X,V)\label{Yu4}\\
&=H(Y)-H(Z|V)\label{Yu5}\\
&\leq 1-H(Z|V)\label{Yu6}
\end{align}
where (\ref{Yu6}) follows by the fact that the entropy of a binary
random variable is upper bounded by $1$. Next, consider the other
cut,
\begin{align}
I(X;Y|T)&=H(Y|T)-H(Y|X,T)\label{Yu7}\\
&= H(Y|T)-H(Z|T)\label{Yu8}\\
&\leq 1-H(Z|T)\label{Yu9}
\end{align}

We note that (\ref{Yu6}) and (\ref{Yu9}) are achieved with
equality for a uniform $X$. Moreover, from (\ref{Yu6}) and
(\ref{Yu9}), it can be observed that the bound $I(X;Y|T)$ is
redundant since $V\rightarrow T\rightarrow Z$ implies $H(Z|T)\leq
H(Z|V)$.  Hence, our upper bound reduces to
\begin{align}
\mathcal{UB}&=\max_{p(v|t):I(T;V)\leq R_{0}} 1-H(Z|V)\label{Yu10}
\end{align}
We should remark that the converse obtained in \cite{YuRelay:2007}
for this channel utilized the modulo-additive nature of the
channel. For such a channel, a uniform distribution on $X$ makes
the channel output $Y$ independent of noise $Z$, thereby making
the proceedings in the converse easier. Our upper bound does not
rely on the nature of the channel and holds for any $p(y|x,t)$.

We have thus shown that for all the cases where the capacity is
established, our bound is tight. To illustrate the usefulness of our
bound, we will consider a channel which does not fall into any of
these classes.

\section{Capacity Result for a Symmetric Binary Erasure Channel with Two States}
We will show that for a particular symmetric binary input erasure
channel with two states, our upper bound yields the capacity which
turns out to be strictly less than the cut-set bound. The state
$T$ is binary with $\mbox{Pr}(T=0)=\alpha$. The
channel input $X$ is binary and channel output $Y$ is ternary. For
channel states $T=0,1$, the transition matrices $p(y|x,t)$ are
given as (Figure $2$),
\begin{align}
W_{0}&=\left[
  \begin{array}{ccc}
    0 & 1-\epsilon & \epsilon \\
    \epsilon & 1-\epsilon & 0 \\
  \end{array}
\right]\hspace{0.2in} W_{1}=\left[
  \begin{array}{ccc}
    \epsilon & 1-\epsilon & 0 \\
    0 & 1-\epsilon & \epsilon \\
  \end{array}
\right]\nonumber
\end{align}

It should be noted that this class of channels does not fall into
the class of channels considered in \cite{KimRelay:2008} since $T$
cannot be obtained as a deterministic function of $X$ and $Y$.
Moreover, the channel output $Y$ cannot be expressed in the form
as $Y=X\oplus Z$, for some $p(t|z)$, where $\oplus$ is modulo-$2$
addition, since the cardinality of $Y$ is different from the
cardinality of $X$. Hence, the converse technique developed in
\cite{YuRelay:2007} for modulo-additive relay channels does not
apply to this channel. However, our upper bound holds for any
$p(y|x,t)$. We begin by evaluating the achievable rates given by
the CAF scheme,
\begin{align}
\mathcal{C}\geq &\sup I(X;Y|V)\nonumber\\
& \text{s.t. } I(T;V|Y)\leq R_{0}\nonumber\\
& \text{for some } p(x,t,v)=p(x)p(t)p(v|t)\label{caf}
\end{align}
Throughout this paper, we denote the entropy function as
\begin{align}
h^{(k)}(s_{1},\ldots,s_{k})=-\sum_{i=1}^{k}s_{i}\text{log}(s_{i})\label{e2}
\end{align}
where $s_{i}\geq 0$, $i=1,\ldots,k$ and $\sum_{i}s_{i}=1$. We will
denote the binary entropy function as $h(s)$. We first define
$\mbox{Pr}(X=0)=p$ and obtain the involved probabilities,
\begin{align}
p(Y=0)&=\epsilon(\alpha*p)\label{e3}\\
p(Y=1)&=1-\epsilon\label{e4}\\
p(Y=2)&=\epsilon(1-\alpha*p)\label{e5}
\end{align}
and
\begin{align}
p(Y=0|T=0)&=\epsilon (1-p)\label{e6}\\
p(Y=1|T=0)&=1-\epsilon\label{e7}\\
p(Y=2|T=0)&=\epsilon p\label{e8}
\end{align}
and
\begin{align}
p(Y=0|T=1)&=\epsilon p\label{e9}\\
p(Y=1|T=1)&=1-\epsilon\label{e10}\\
p(Y=2|T=1)&=\epsilon (1-p)\label{e11}
\end{align}
where we have defined
\begin{align}
a*b=a(1-b)+b(1-a)
\end{align}
Furthermore, we also note the following inequality,
\begin{figure}[t]
  \centering
  \centerline{\epsfig{figure=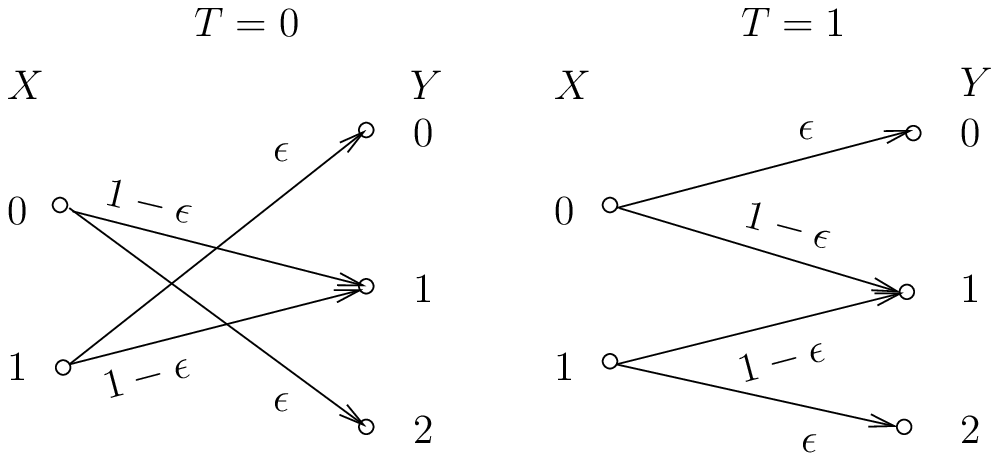,width=8cm}}
  { Figure $2$: A symmetric binary erasure channel with two states.}\medskip
  \vspace{-0.3in}
\end{figure}
\begin{align}
h^{(3)}(a,b,c)&=\frac{1}{2}h^{(3)}(a,b,c)+\frac{1}{2}h^{(3)}(c,b,a)\label{e12}\\
&\leq h^{(3)}\left(\frac{a+c}{2},b,\frac{a+c}{2}\right)\label{e13}\\
&=h(b)+1-b\label{e14}
\end{align}
Using this fact, we have
\begin{align}
H(Y)&=h^{(3)}\left(\epsilon(\alpha*p),1-\epsilon,\epsilon(1-\alpha*p)\right)\label{e15}\\
&\leq h(\epsilon)+\epsilon\label{e16}
\end{align}
Also, a uniform distribution on $X$, yields the maximum entropy
for $Y$, and makes $Y$ and $T$ independent. Note that the maximum
entropy of $Y$ in this case is $h(\epsilon)+\epsilon$ which is
strictly less than $\mbox{log}(3)$ for all $\epsilon \in [0,1]$.
Hence, for a uniform $X$, we have
\begin{align}
H(Y|V)&=H(Y)\label{e17}\\
&=h(\epsilon)+\epsilon\label{e18}
\end{align}
We also define,
\begin{align}
\eta_{v}&=\mbox{Pr}(T=1|V=v), \hspace{0.05in}
v=1,\ldots,|\mathcal{V}|\label{e19}
\end{align}
Using this definition, we can write $H(Y|X,V)$ for any
distribution $p(x)$ on $X$ as follows,
\begin{align}
H(Y|X,V)&=\sum_{v}p(v)\sum_{x}p(x)H(Y|X=x,V=v)\label{e20}\\
&=\sum_{v}p(v)h^{(3)}\left(\eta_{v}\epsilon,1-\epsilon,(1-\eta_{v})\epsilon\right)\label{e21}\\
&=H(U|V)\label{e22}
\end{align}
where we have defined a random variable $U$ with $|\mathcal{U}|=3$
and $p(u|t)$, expressed as a stochastic matrix $W$ which is given
as
\begin{align}
W&=\left(%
\begin{array}{ccc}
  \epsilon & 1-\epsilon & 0 \\
  0 & 1-\epsilon & \epsilon \\
\end{array}%
\right)\label{WU}
\end{align}
Thus, $H(Y|X,V)$ is invariant to the distribution of $X$.
Moreover, by construction, the random variables $(T,U,V)$ satisfy
the Markov chain $V\rightarrow T \rightarrow U$.

We now return to the evaluation of the rates given by the CAF
scheme given in (\ref{caf}). Using (\ref{e18}) and (\ref{e22}), we
have for a uniform distribution on $X$,
\begin{align}
I(X;Y|V)&=H(Y|V)-H(Y|X,V)\label{e23}\\
&=h(\epsilon)+\epsilon-H(U|V)\label{e24}
\end{align}
Furthermore, for uniform $X$, we have $I(T;V|Y)=I(T;V)$, thus the
constraint in (\ref{caf}) simplifies to $I(T;V)\leq R_{0}$. For
simplicity, define the set
\begin{align}
\mathcal{L}(\gamma)&=\{p(v|t): H(T|V)\geq \gamma;
\hspace{0.05in}V\rightarrow T\rightarrow U\}\label{e25}
\end{align}
Using (\ref{e24}) and (\ref{e25}), we obtain a lower bound on the
capacity as
\begin{align}
\mathcal{C}\geq h(\epsilon)+\epsilon-\inf_{p(v|t)\in
\mathcal{L}(h(\alpha)-R_{0})}H(U|V)\label{e26}
\end{align}

We now evaluate our upper bound. Using the following fact,
\begin{align}
\min(I(X,V;Y),I(X;Y|T))\leq I(X,V;Y)
\end{align}
we obtain a weaker version
of our upper bound in (\ref{UB}) as
\begin{align}
\mathcal{C}&\leq \sup I(X,V;Y) \label{e27}\\
&= \sup (H(Y)-H(Y|X,V))\label{e28}\\
&\leq  \sup (h(\epsilon)+\epsilon-H(Y|X,V))\label{e29}\\
&= h(\epsilon)+\epsilon-\inf H(Y|X,V)\label{e30}\\
&= h(\epsilon)+\epsilon-\inf_{p(v|t)\in
\mathcal{L}(h(\alpha)-R_{0})} H(U|V)\label{e31}
\end{align}
where (\ref{e29}) follows from (\ref{e16}), and the $\sup$ in
(\ref{e27})-(\ref{e29}) is taken over all $p(x)$ and those
$p(v|t)$ which satisfy $I(T;V)\leq R_{0}$.

Hence, from (\ref{e26}) and (\ref{e31}), the capacity is given by
\begin{align}
\mathcal{C}&= h(\epsilon)+\epsilon-\inf_{p(v|t)\in
\mathcal{L}(h(\alpha)-R_{0})}H(U|V)\label{e32}
\end{align}
We will now explicitly evaluate the capacity expression obtained
in (\ref{e32}) and compare it with the cut-set bound. For this
purpose, we need a result on the conditional entropy of dependent
random variables \cite{Witsenhausen:1975}. Let $T,U$ be a pair of
dependent random variables with a joint distribution $p(t,u)$. For
$0\leq \gamma \leq H(T)$, define the function $G(\gamma)$ as the
infimum of $H(U|V)$, with respect to all discrete random variables
$V$ such that $H(T|V)=\gamma$ and the random variables $V$ and $U$
are conditionally independent given $T$. For the case when $T$ is
binary and $p(u|t)$, expressed as a stochastic matrix $W$, takes
the form in (\ref{WU}), we have from \cite{Witsenhausen:1975},
\begin{align} G(\gamma)&=\inf_{p(v|t) \in \mathcal{L}(\gamma)}H(U|V)\\
&=h(\epsilon)+\epsilon \gamma \label{e33}
\end{align}

We will use this result from \cite{Witsenhausen:1975} in
explicitly evaluating the capacity in (\ref{e32}). First note
that, if $R_{0}\geq h(\alpha)$, then
\begin{align}
G(h(\alpha)-R_{0})&=G(0)=h(\epsilon)\label{e34}
\end{align}
whereas, if $R_{0}<h(\alpha)$, then
\begin{align}
G(h(\alpha)-R_{0})=h(\epsilon)+\epsilon(h(\alpha)-R_{0})\label{e35}
\end{align}
Using (\ref{e34}) and (\ref{e35}), the capacity expression in
(\ref{e32}) evaluates to,
\begin{align}
\mathcal{C}(R_{0})&=\left\{%
\begin{array}{ll}
    \epsilon, & \hbox{$R_{0}\geq h(\alpha)$} \\
    \epsilon(1-h(\alpha))+\epsilon R_{0}, & \hbox{$R_{0}<h(\alpha)$} \\
\end{array}%
\right. \label{e36}
\end{align}
which can be written in a compact form as,
\begin{align}
\mathcal{C}(R_{0})&=\min(\epsilon(1-h(\alpha))+\epsilon
R_{0},\epsilon)\label{e37}
\end{align}
The cut-set bound is obtained by evaluating (\ref{cutset}) for the
channel in consideration. Evaluation of the cut-set bound is
straightforward by noting that $I(X;Y)$ and $I(X;Y|T)$ are both
maximized by a uniform $p(x)$. For a uniform distribution on $X$,
we have the following equalities,
\begin{align}
I(X;Y)&=\epsilon(1-h(\alpha))\\
I(X;Y|T)&=\epsilon
\end{align}
Hence, the cut-set bound is given as,
\begin{align}
\mathcal{CS}(R_{0})&=\min(\epsilon(1-h(\alpha))+R_{0},\epsilon)\label{e38}
\end{align}
The difference between the capacity and the cut-set bound is
evident from the first term in the $\min$ operation, i.e., the
capacity expression in (\ref{e37}) has an $\epsilon R_{0}$
appearing in the minimum, as opposed to $R_{0}$ appearing in the
cut-set bound at the corresponding place in (\ref{e38}). The
cut-set bound and the capacity are shown in Figure $3$ as
functions of $R_{0}$ for $\alpha=0.3$ and $\epsilon=0.4$.

In conclusion, for this channel which does not fall into the
classes of channels studied in \cite{KimRelay:2008} and
\cite{YuRelay:2007}, our upper bound equals the CAF achievable
rate, thus yielding the capacity, which is strictly less than the
cut-set bound for $R_{0}<h(\alpha)$.

\section{A Channel with Binary Multiplicative State and Binary Additive Noise}
We will evaluate our upper bound and compare it with the cut-set
bound for the case when $X$, $T$ and $N$ are binary and the channel
is given as,
\begin{align}
Y&=TX+N\label{eg1}
\end{align}
The channel output $Y$ takes values in the set $\{0,1,2\}$. The
random variables $T$ and $N$ are distributed as $T\sim
\mbox{Ber}(\alpha)$ and $N\sim \mbox{Ber}(\delta)$. This relay
channel does not fall into the subclass of channels considered in
\cite{KimRelay:2008}. Moreover, the converse obtained in
\cite{YuRelay:2007} does not apply to this channel since the
output cannot be written as a modulo-sum.

To evaluate our upper bound, let us define
\begin{align}
\mbox{Pr}(X=1)&=p\label{eg3}\\
\mbox{Pr}(T=1)&=\alpha \label{eg4}\\
\mbox{Pr}(N=1)&=\delta \label{eg5}
\end{align}
We then obtain $H(Y)$ as follows
\begin{align}
H(Y)&=h^{(3)}(P_{Y}(0),P_{Y}(1),P_{Y}(2))\label{eg6}
\end{align}
where
\begin{align}
P_{Y}(0)&=p(1-\alpha)(1-\delta)+(1-p)(1-\delta)\label{eg7}\\
P_{Y}(1)&=(1-p)\delta+p[(1-\alpha)\delta+\alpha(1-\delta)]\label{eg8}\\
P_{Y}(2)&=p\alpha\delta\label{eg9}
\end{align}
and $H(Y|X)$ is obtained as,
\begin{align}
H(Y|X)&=(1-p)H(N)+pH(T+N)\label{eg10}\\
&=(1-p)h(\delta)+p
h^{(3)}((1-\alpha)(1-\delta),\alpha*\delta,\alpha\delta)\label{eg11}
\end{align}
The broadcast cut is obtained as,
\begin{align}
I(X;Y|T)=& \hspace{0.05in}H(Y|T)-H(Y|X,T)\label{eg12}\\
=& \hspace{0.05in}(1-\alpha)h(\delta)+\alpha
h^{(3)}((1-p)(1-\delta),p*\delta,p\delta)\nonumber\\
&\hspace{0.03in}-h(\delta)\label{eg13}
\end{align}
The cut-set bound is given by,
\begin{align}
\mathcal{CS}&=\max_{p}\min{\{I(X;Y)+R_{0},I(X;Y|T)\}}\label{eg14}
\end{align}
\begin{figure}[t]
  \centering
  \centerline{\epsfig{figure=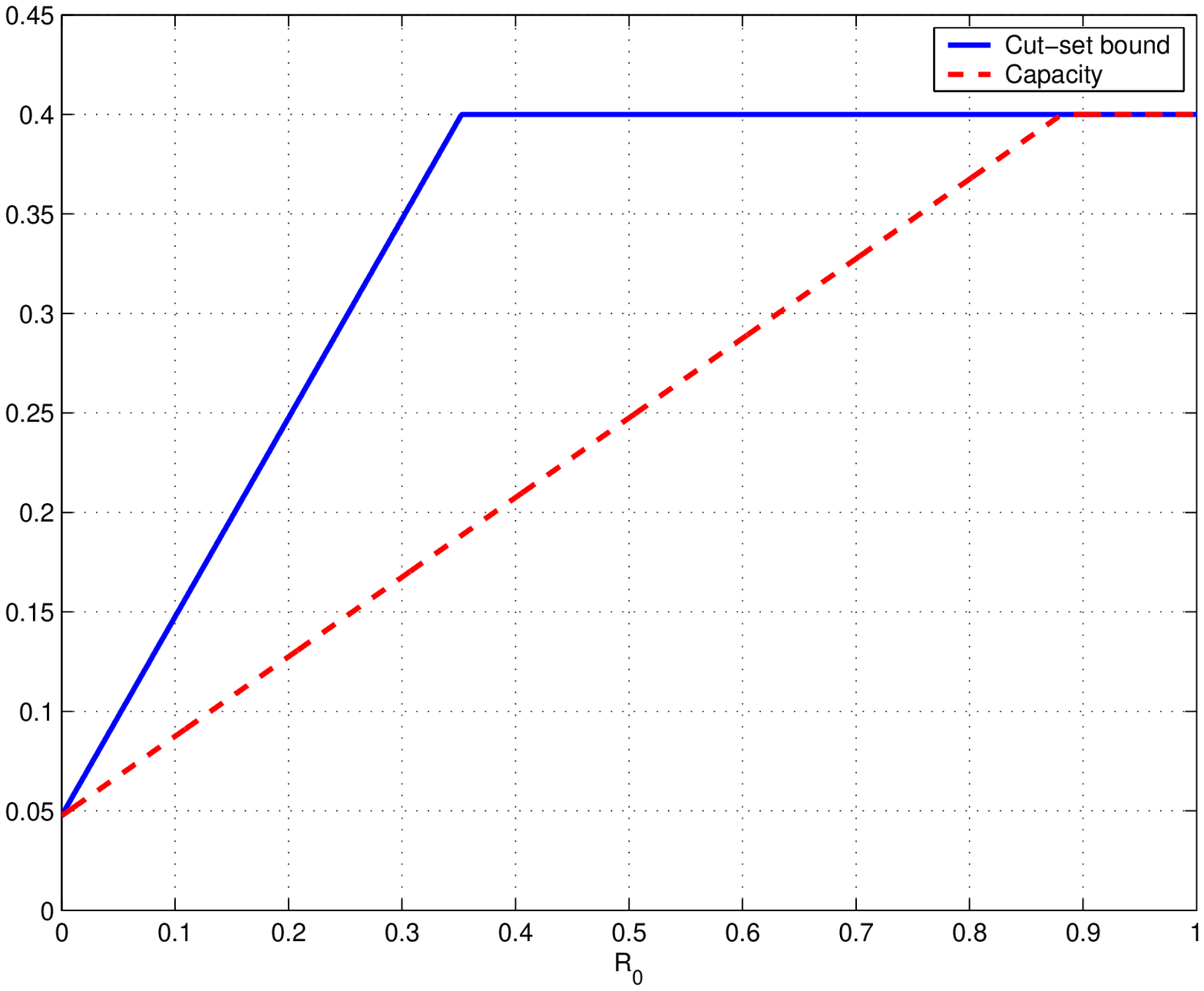,width=8cm}}
  { Figure $3$: Capacity of the binary symmetric erasure channel for $\alpha=0.3$ and $\epsilon=0.4$.}\medskip
\end{figure}

We now evaluate our bound by first considering,
\begin{align}
I(X,V;Y)&=H(Y)-H(Y|X,V)\label{eg15}
\end{align}
We have already evaluated $H(Y)$ in (\ref{eg6}). Consider
$H(Y|X,V)$:
\begin{align}
H(Y|X,V)=&\sum_{(x,v)}P_{X}(x)P_{V}(v)H(Y|X=x,V=v)\label{eg16}\\
=&\sum_{v}P_{V}(v)\big[(1-p)H(Y|X=0,V=v)\nonumber\\
& \hspace{0.57in}+pH(Y|X=1,V=v)\big]\label{eg17}\\
=&\sum_{v}P_{V}(v)\big[(1-p)H(N)\nonumber\\&\hspace{0.57in}+pH(T+N|V=v)\big]\label{eg18}\\
=&\sum_{v}P_{V}(v)\big[(1-p)h(\delta)+pH(T+N|V=v)\big]\label{eg19}\\
=&\sum_{v}P_{V}(v)\big[(1-p)h(\delta)+pH(W|V=v)\big]\label{eg20}
\end{align}
\begin{align}
=&\hspace{0.05in}(1-p)h(\delta)+pH(W|V)\label{eg21}
\end{align}
where we have defined another random variable $W$ as follows,
\begin{align}
W=T+N
\end{align}
We are interested in lower bounding $H(W|V)$. We also know that any
permissible conditional distribution $p(v|t)$ satisfies the
constraint $I(T;V)\leq R_{0}$. Using this, we also have the
following,
\begin{align}
H(T|V)&\geq h(\alpha)-R_{0}\label{eg22}
\end{align}
Let us also define,
\begin{align}
P_{T|V}(T=1|V=v)&=\eta_{v}, \hspace{0.05in} v\in
1,\ldots,|\mathcal{V}|\label{eg23}
\end{align}
We now return to calculating $H(W|V)$
\begin{align}
\mbox{Pr}(W=w|V=v)=&\sum_{t}P_{T|V}(t|v)P_{W|T,V}(w|t,v)\label{eg25}\\
=&\hspace{0.05in}(1-\eta_{v})P(w|T=0,V=v)\nonumber\\&+\eta_{v}P(w|T=1,V=v)\label{eg26}
\end{align}
Since the random variable $W$ takes values in the set $\{0,1,2\}$,
we obtain,
\begin{align}
\mbox{Pr}(W=0|V=v)&=(1-\eta_{v})(1-\delta) \label{eg27}\\
\mbox{Pr}(W=1|V=v)&= \eta_{v}*\delta \label{eg28}\\
\mbox{Pr}(W=2|V=v)&= \eta_{v}\delta \label{eg29}
\end{align}
We finally obtain,
\begin{align}
H(W|V)=\sum_{v}P_{V}(v)h^{(3)}((1-\eta_{v})(1-\delta),\eta_{v}*\delta,\eta_{v}\delta)\label{eg30}
\end{align}
For the special case when the additive noise is $N\sim
\mbox{Ber}(1/2)$, the above expression simplifies to
\begin{align}
H(W|V)&=\sum_{v}P_{V}(v)h^{(3)}\left(\frac{(1-\eta_{v})}{2},\frac{1}{2},\frac{\eta_{v}}{2}\right)\label{eg31}\\
&=\sum_{v}P_{V}(v)\Bigg(\frac{1}{2}h(\eta_{v})+1\Bigg)\label{eg32}\\
&=\frac{1}{2}H(T|V)+1\label{eg33}\\
&\geq \frac{1}{2}\left(h(\alpha)-R_{0}\right)+1\label{eg34}
\end{align}
where (\ref{eg34}) follows from (\ref{eg22}). Substituting
(\ref{eg34}) in (\ref{eg21}) we obtain
\begin{align}
H(Y|X,V)&=(1-p)h(\delta)+pH(W|V)\label{eg35}\\
&\geq
(1-p)h(\delta)+p\Bigg(\frac{1}{2}\left(h(\alpha)-R_{0}\right)+1\Bigg)\label{eg36}
\end{align}
Continuing from (\ref{eg15}), we obtain an upper bound on
$I(X,V;Y)$ as follows,
\begin{align}
I(X,V;Y)&=H(Y)-H(Y|X,V)\label{eg37}\\
&\leq H(Y)-1-\frac{p}{2}(h(\alpha)-R_{0})\label{eg38}
\end{align}
Moreover, the first term appearing in the cut-set bound simplifies
to
\begin{align}
I(X;Y)+R_{0}&=H(Y)-H(Y|X)+R_{0}\label{eg39}\\
&=H(Y)-1-\frac{p}{2}h(\alpha)+R_{0}\label{eg40}
\end{align}

We thus obtain our upper bound as,
\begin{align}
\mathcal{UB}&=\max_{p\in[0,1]}\min\Big[H(Y)-1-\frac{p}{2}h(\alpha)+pR_{0},I(X;Y|T)\Big]\label{eg41}
\end{align}
whereas the cut-set bound is,
\begin{align}
\mathcal{CS}&=\max_{p\in
[0,1]}\min\Big[H(Y)-1-\frac{p}{2}h(\alpha)+R_{0},I(X;Y|T)\Big]\label{eg42}
\end{align}
The difference between the cut-set bound and our upper bound is
evident from the first term in the $\min$ operation, i.e., our
upper bound has a $pR_{0}$ term in (\ref{eg41}), as opposed to
$R_{0}$ at the corresponding place in (\ref{eg42}).

Both these bounds along with the CAF rate are illustrated in
Figure $4$ as a function of $R_{0}$ for the case when $\alpha=1/2$
and $\delta=1/2$. We should remark here that although our bound is
strictly smaller than the cut-set bound for certain values of
$R_{0}$, it is strictly larger than the rates given by the CAF
scheme. Here, the CAF rates are evaluated by restricting $V$ to be
binary, i.e., by considering all conditional distributions
$p(v|t)$, such that, $|\mathcal{V}|=2$. Therefore, the CAF rates
plotted in Figure $4$ are potentially suboptimal and can be
potentially improved upon by increasing the cardinality of $V$.
\section{Discussion}
Let us recall our upper bound obtained in (\ref{UB}),
\begin{align}
\mathcal{UB}&= \sup \min \{I(X,V;Y),I(X;Y|T)\}\nonumber\\
&\hspace{0.2in}\text{s.t. }R_{0}\geq I(T;V)\nonumber\\
&\hspace{0.2in}\text{over }p(x)p(t)p(v|t)\label{UB1}
\end{align}
Using the fact that
\begin{align}
\min(I(X,V;Y),I(X;Y|T))\leq I(X,V;Y)
\end{align}
and observing that
\begin{align}
I(X,V;Y)=I(V;Y)+I(X;Y|V)
\end{align}
it can be noted that
our upper bound in (\ref{UB1}) can be further upper bounded as
\begin{align}
\mathcal{C}\leq& \sup I(V;Y)+I(X;Y|V)\\
&\mbox{s.t. }I(T;V)\leq R_{0}\\
&\mbox{for some }p(x)p(v|t)
\end{align}
On the other hand, the capacity is always lower bounded by the CAF
rate,
\begin{align}
\mathcal{C}\geq& \sup I(X;Y|V)\\
&\mbox{s.t. }I(T;V|Y)\leq R_{0}\\
&\mbox{for some }p(x)p(v|t)
\end{align}
Now using the following fact,
\begin{align}
I(T;V|Y)&=H(V|Y)-H(V|T)\\
&=I(T;V)-I(V;Y)
\end{align}
we can rewrite the CAF lower bound on the capacity as
\begin{align}
\mathcal{C}\geq& \sup I(X;Y|V)\\
&\mbox{s.t. }I(T;V)-I(V;Y)\leq R_{0}\\
&\mbox{for some }p(x)p(v|t)
\end{align}
We can see that the CAF lower bound on the capacity involves
taking a supremum of $I(X;Y|V)$ subject to the constraint
$I(T;V)-I(V;Y)\leq R_{0}$ whereas our upper bound involves taking
a supremum of a larger quantity $I(V;Y)+I(X;Y|V)$ subject to a
stricter constraint $I(T;V)\leq R_{0}$.

Although these two maximization problems are different, for the
class of channels for which capacity was obtained, at the capacity
achieving input distribution $p(x)$, we had $I(V;Y)=0$. Moreover,
the same input distribution $p(x)$ yielded the maximum for both
maximization problems.
 Thus, for the
class of channels considered in Section \textrm{VI}, these two
maximization problems are equivalent. This observation yields a
heuristic explanation as to why we were able to obtain the
capacity results for these classes of channels.
\begin{figure}[t]
  \centering
  \centerline{\epsfig{figure=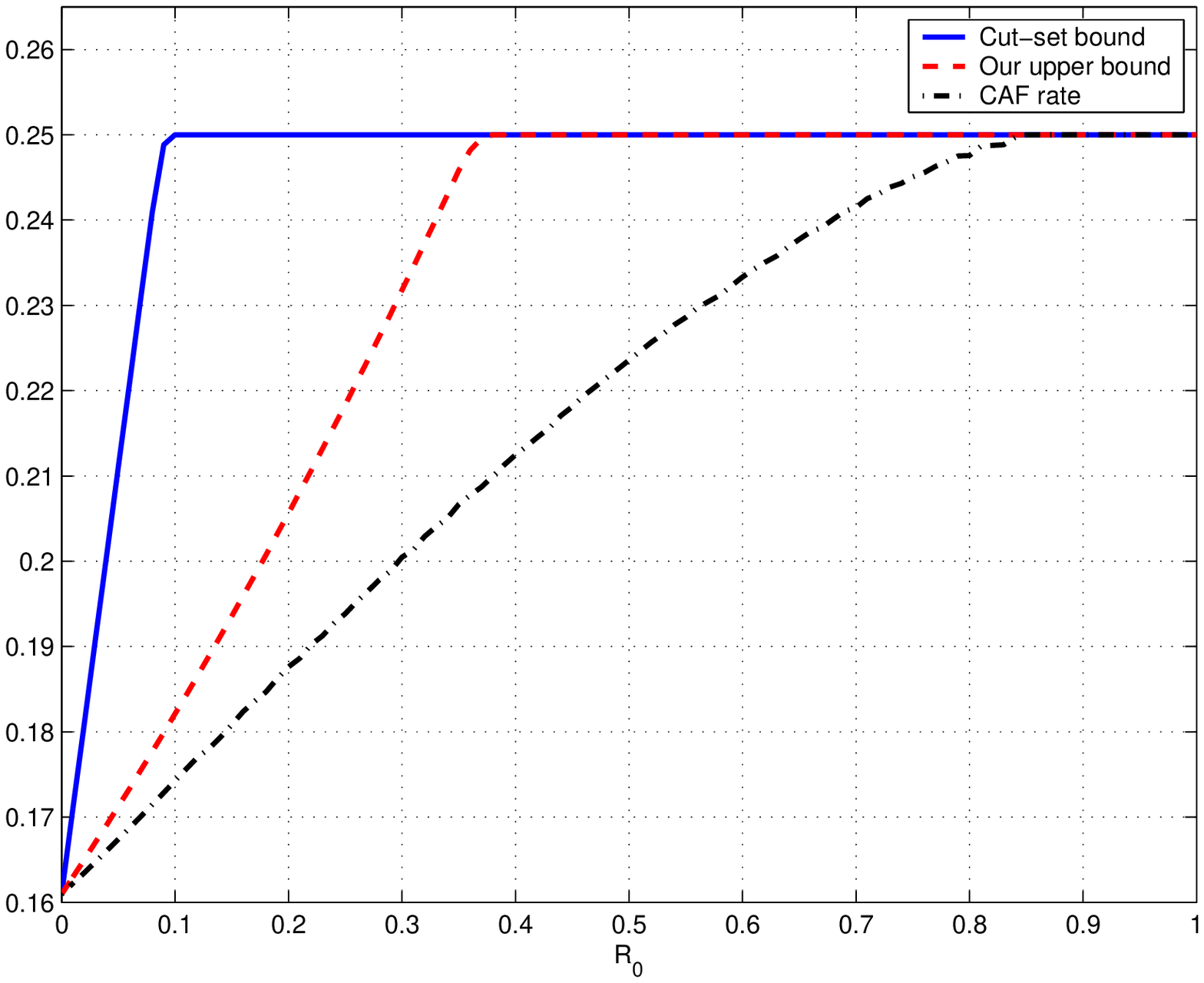,width=8cm}}
  { Figure $4$: Comparison of our upper bound with the cut-set bound when $T\sim \mbox{Ber}(1/2)$ and $N\sim \mbox{Ber}(1/2)$.}\medskip
\end{figure}

\section{A New Lower Bound on Critical $R_{0}$ }

In \cite{CoverRelay:1987}, Cover posed a slightly different
problem regarding the general primitive relay channel. Considering the
capacity as a function of $R_{0}$, i.e., $\mathcal{C}(R_{0})$,
first observe the following facts,
\begin{align}
\mathcal{C}(0)&=\sup_{p(x)}I(X;Y)\label{crit1}\\
\mathcal{C}(\infty)&=\sup_{p(x)}I(X;Y|T)\label{crit2}
\end{align}
Moreover, $\mathcal{C}(R_{0})$ is a nondecreasing function of
$R_{0}$. Cover posed the following question in
\cite{CoverRelay:1987}: what is the smallest value of $R_{0}$, say
$R_{0}^{*}$, for which
$\mathcal{C}(R_{0}^{*})=\mathcal{C}(\infty)$? As an application of
our upper bound, we implicitly provide a new lower bound on
$R_{0}^{*}$ for the class of primitive relay channels studied in this paper.

For the class of channels considered in Section \textrm{VI}, we
obtained the capacity. As a consequence, we can explicitly
characterize $R_{0}^{*}$ for this class of channels as
$h(\alpha)$. Furthermore, for the class of channels considered in
Section \textrm{VII}, our upper bound on the capacity yields an
improved lower bound on $R_{0}^{*}$ than the one provided by the
cut-set bound, which is clearly evident in Figure $4$.

\section{Conclusions}
We obtained a new upper bound for a class of primitive relay
channels. The primitive relay channel studied in this paper can
also be considered as a state-dependent discrete memoryless
channel, with rate-limited state information available at the
receiver and no state information available at the transmitter.

Using our upper bound, we first recover all previously known
capacity results for such channels. Furthermore, we explicitly
characterize the capacity of a new subclass of these primitive
relay channels which does not overlap with the classes previously
studied in \cite{KimRelay:2008},\cite{YuRelay:2007}. In
particular, for this class of channels, it is assumed that there
are two channel states, and for each channel state, there is an
erasure channel from $X$ to $Y$. We show that the capacity for
such channels is strictly smaller than the cut-set bound for
certain values of $R_{0}$. This capacity result  validates a
conjecture due to Ahlswede and Han \cite{AhlswedeHan:1983} for
this class of channels.

Moreover, we also evaluated our upper bound for a case where
$Y=TX+N$, where $T,X$ and $N$ are binary. This channel does not
fall into any of the classes studied in
\cite{KimRelay:2008},\cite{YuRelay:2007} and neither does it fall
into the aforementioned class of channels. We show that our upper
bound strictly improves upon the cut-set bound for certain values
of $R_{0}$, although, our upper bound is strictly larger than the
rates yielded by a potentially suboptimal evaluation of the CAF
scheme.

\bibliography{allerton08}

\begin{thebibliography}{10}

\bibitem{KimRelay:2008}
Y-H. Kim.
\newblock Capacity of a class of deterministic relay channels.
\newblock {\em IEEE Trans. on Information Theory}, 54(3):1328--1329, Mar. 2008.

\bibitem{YuRelay:2007}
M.~Aleksic, P.~Razaghi, and W.~Yu.
\newblock Capacity of a class of modulo-sum relay channels".
\newblock {\em \emph{Submitted to }IEEE Trans. on Information Theory}, 2007.

\bibitem{Cover:1979}
T.~M. Cover and A.~El~Gamal.
\newblock Capacity theorems for the relay channel.
\newblock {\em IEEE Trans. on Information Theory}, 25(5):572--584, September
  1979.

\bibitem{Zhang:1988}
Z.~Zhang.
\newblock Partial converse for the relay channel.
\newblock {\em IEEE Trans. on Information Theory}, 34(5):1106--1110, September
  1988.

\bibitem{KimAllerton:2007}
Y-H. Kim.
\newblock Coding techniques for primitive relay channels.
\newblock {\em Proc. Annual Allerton Conference on Communication, Control and
  Computing}, pages 129--135, 2007.

\bibitem{Cover:book}
T.~M. Cover and J.~A. Thomas.
\newblock {\em Elements of Information Theory}.
\newblock New York:Wiley, 1991.

\bibitem{HeegardElGamal:1983}
C.~Heegard and A.~El Gamal.
\newblock On the capacity of computer memory with defects.
\newblock {\em IEEE Trans. on Information Theory}, 29(5):731--739, Sep. 1983.

\bibitem{AhlswedeHan:1983}
R.~Ahlswede and T.~S. Han.
\newblock On source coding with side information via a multiple-access channel
  and related problems in multi-user information theory.
\newblock {\em IEEE Trans. on Information Theory}, 29(3):396--412, May 1983.

\bibitem{Witsenhausen:1975}
H.~S. Witsenhausen and A.~D. Wyner.
\newblock A conditional entropy bound for a pair of discrete random variables.
\newblock {\em IEEE Trans. on Information Theory}, 21(5):493--501, September
  1975.

\bibitem{CoverRelay:1987}
T.~M. Cover.
\newblock The capacity of the relay channel.
\newblock In {\em Open Problems in Communication and Computation}, pages
  72--73, 1987.

\end{thebibliography}
\bibliographystyle{unsrt}
\end{document}